\documentclass[3p,times,procedia]{elsarticle}
\usepackage{nupha_ecrc}
\usepackage{graphicx}
\usepackage{bm,amsmath,amssymb}
\usepackage[figuresright]{rotating}

\newcommand{\beq}{\begin{equation}}
\newcommand{\eeq}{\end{equation}}

\newcommand{\bp}{\bm{p}}
\newcommand{\bx}{\bm{x}}

\newcommand{\mcal}{\mathcal}
\newcommand{\rmd}{{\rm d}}
\newcommand{\rme}{{\rm e}}
\newcommand{\eqn}[1]{Eq.~(\ref{#1})}

\newcommand{\tbr}{t_{\rm br}}
\newcommand{\trel}{t_{\rm rel}}

\long\def\comment#1{ }

\newcommand{\x}{{\rm x}}

\newcommand{\be}{\begin{equation}}
\newcommand{\ee}{\end{equation}}
\newcommand{\bea}{\begin{eqnarray}}
\newcommand{\eea}{\end{eqnarray}}
\newcommand{\f}{\frac}

\newcommand{\bra}{\langle}
\newcommand{\ket}{\rangle}

\volume{00}

\firstpage{1}

\journalname{Nuclear Physics A}

\runauth{E. Iancu and B. Wu}

\jid{nupha}

\jnltitlelogo{Nuclear Physics A}

\usepackage{amssymb}
\usepackage[figuresright]{rotating}

\begin{document}

\begin{frontmatter}



\dochead{}
%% e.g. \dochead{17th International Conference on Dynamical Processes in Excited States of Solids}

\title{Thermalization of mini-jets in a quark-gluon plasma}

\author{Edmond Iancu}
\ead{edmond.iancu@cea.fr}
  \address{Institut de Physique Th\'{e}orique, CEA Saclay, CNRS UMR 3681, F-91191 Gif-sur-Yvette, France}
    
\author{Bin Wu\corref{cor1}}
\ead{bin.wu.phys@gmail.com}

  \cortext[cor1]{Corresponding author}
 \address{Institut de Physique Th\'{e}orique, CEA Saclay, CNRS UMR 3681, F-91191 Gif-sur-Yvette, France\\
 Department of Physics, The Ohio State University, Columbus
, OH 43210, USA}

\begin{abstract}

We present the complete physical picture for the evolution of a high-energy jet propagating through a weakly-coupled quark-gluon plasma (QGP) by analytical and numerical investigation of thermalization of the soft components of the jet. Our results support the following physical picture: the leading particle emits a significant number of mini-jets which promptly evolve via multiple branching and thus degrade into a myriad of soft gluons, with energies of the order of the medium temperature $T$. Via elastic collisions with the medium constituents, these soft gluons relax to local thermal equilibrium with the plasma over a time scale which is considerably shorter than the typical lifetime of the mini-jet. The thermalized gluons form a tail which lags behind the hard components of the jet. Together with the background QGP, they behave hydrodynamically.
\end{abstract}


\end{frontmatter}

% \linenumbers

\section{Introduction}

A striking imbalance in dijet transverse momentum has been observed in central lead-lead collisions at the LHC\cite{Aad:2010bu,Chatrchyan:2011sx}. Detailed data analysis indicates that the momentum imbalance is accompanied by a
softening of the fragmentation pattern of the second most energetic away-side jet\cite{Chatrchyan:2011sx}. One natural theoretical interpretation of such a so-called jet quenching phenomenon is that the away-side jet has enough time to degrade into soft gluons due to medium-induced multiple branching. The emergent new picture for the in-medium jet evolution is as follows: the energy of the hard components of the jet is efficiently transmitted, 
via multiple, quasi-democratic, branchings,
to a multitude of comparatively soft gluons\cite{Baier:2000sb, Blaizot:2012fh, Blaizot:2013hx,Fister:2014zxa,Kurkela:2014tla,Apolinario:2014csa}, which relax to local thermal equilibrium with the plasma over a time scale considerably shorter than the branching time of the jet\cite{Iancu:2015uja}. In particular, Ref.~\cite{Iancu:2015uja} presented for the first time the picture of this
evolution in longitudinal phase-space, with the longitudinal axis defined as the
direction of propagation of the leading particle. This picture
will be succinctly summarized in what follows.
\label{sec:Intro}

\section{Kinetic theory for jet evolution}
\label{sec:kin}

We study the parton distribution produced by a high-energy jet propagating
through a weakly-coupled quark-gluon plasma in thermal equilibrium at temperature $T$. 
The leading particle (LP) which initiates the jet has a high
energy $E\gg T$ and a comparatively small virtuality. 
We concentrate on the {\em medium-induced} evolution, as
triggered by the collisions between the partons from the jet and the constituents of the medium.
One can distinguish between two types of collisions:

\texttt{(i)} elastic, $2\! \to\! 2$, collisions, which entail energy and momentum transfer between
the jet and the medium;

\texttt{(ii)} inelastic collisions, like $2\to 3$ or, more generally, 1+(many)$\to$ 2+(many), in which
a parton from the jet undergoes a $1\to 2$ branching.

At weak coupling, such processes can be described by a kinetic equation for the gluon
distribution \cite{Baier:2000sb,Arnold:2002zm},
\beq\label{generaleq}
\left(\frac{\partial}{\partial t}+{\bm v}\cdot \nabla_{\bm x} \right)f(t,\bx,\bp)\, =\, \mcal{C}_{\rm el}[f]+
\mcal{C}_{\rm br}[f]\,.\eeq
Here, $f(t,\bx,\bp)$ is the gluon occupation number, %in the 3-D phase-space
${\bm v}=\bp/p$ with $p\equiv |\bp|$ is the gluon velocity. In diffusion approximation (see \cite{Hong:2010at, Blaizot:2014jna} and references therein), the elastic collision term takes the form:
\beq\label{CFP}
\mcal{C}_{\rm el}[f]\,\simeq\,
\frac{1}{4}\,\hat{q}\,\nabla_{\bp}\cdot\left[ \left( \nabla_{\bp} + \frac{{\bm v}}{T}\right) f\right]\,,
\eeq
with $\hat q\!\sim\! \alpha_s^2 T^3\ln(1/\alpha_s)$ 
the jet quenching parameter. One can easily check that \eqn{CFP} admits the
Maxwell-Boltzmann thermal distribution $f_{\rm eq}\propto \rme^{-p/T}$  as a fixed point, which shall be reached for a gluon with $p\sim T$ within a time interval of order
\beq\label{trel}
\trel\equiv \frac{4T^2}{\hat{q}}\,\sim\, \frac{1}{\alpha_s^2 T\ln(1/\alpha_s)}\,.
\eeq
The inelastic collision integral $\mcal{C}_{\rm br}[f]$ including only medium-induced gluon branching in the LPM regime is of the form\cite{Baier:2000sb}
\bea\label{Cbr}
\mcal{C}_{\rm br}[f]\,\simeq\,
\f{1}{\tbr(p)}\int_0^1 \rmd x\,  {\cal K}(x)\left[\f{1}{x^{\f{5}{2}}}f\left(t,\bx,{\f{{\bm p}}{x}}\right) - \f{1}{2} f(t,\bx,{{\bm p}}) \right]
\eea
with
$\tbr(p)= \frac{1}{\alpha_s}\sqrt{\frac{p}
 {\hat q}}$ and the splitting kernel\cite{Baier:1996kr,Zakharov:1996fv} 
\be\label{Kbr}
{\cal K}(x)\equiv\frac{[1-x(1-x)]^{\frac{5}{2}}}{[x(1-x)]^{\frac{3}{2}}}\qquad\text{with}\qquad 0\le x \le 1.
\ee

During most stages of the branching process, the cascade is built with relatively 
hard gluons, which are nearly collinear with the LP:
$p_z\gg p_\perp$. It therefore makes sense to focus on the {\em longitudinal} dynamics,
as obtained after integrating out the transverse phase-space. 
This motivates
the following, relatively simple, kinetic equation, for the
{\em longitudinal gluon distribution} $f_\ell(t,z,p_z)\equiv
\int{\rmd^2\bx_\perp \rmd^2\bp_\perp}\,f(t,\bx,\bp)$ \cite{Iancu:2015uja} :
\begin{align}\label{eqL}
&\left(\partial_t+v\partial_z\right)f_\ell(t,z,p)
 = \frac{\hat{q}}{4}{\partial_p}\left[\left(\partial_p + \frac{v}{T}\right) f_\ell(t,z,p) 
 \right]\\*[0.3cm]
&
+\frac{1}{\tbr(p)}\int\limits_r \rmd x\,  {\cal K}(x)\left[\frac{1}{\sqrt{x}}\,
f_\ell\left(t,z,\frac{p}{x}\right) - \frac{1}{2} f_\ell(t,z,p) \right]. \nonumber
\end{align}
Here, $p\equiv p_z$, $v\equiv p/|p|$
and the subscript $r$ on the integral over $x$ means that the branching process
is cut off at the soft scale $p=T$.
%, to mimic non-linear effects like
%gluon recombination which are not explicitly included in this equation.
The two terms within the inelastic collision integral are recognized as the {\em gain}
term and {\em loss} term, respectively.
%in the gain term, a gluon with momentum $p$ is produced via the splitting of a parent gluon with
%momentum $p/x$\,; in the loss term, a gluon with momentum $p$ disappears because it splits.
%The characteristic time scale within the inelastic term is 
 %$\tbr(p)$, in agreement with the previous discussion.
The initial condition reads
\beq\label{init}
f_\ell(t=0,z,p)=\delta(p-E)\delta(z)\,,
\eeq
corresponding to a LP with longitudinal momentum $p=E$ which enters the medium
at $t=0$ and $z=0$. The subscript $l$ of $f_l$ shall be omitted in the following discussions.

\section{The complete physical picture of jet evolution}
\label{sec:sol}

\begin{figure}[t]
\begin{center}
 %\comment
 {\includegraphics[width=0.45\textwidth]{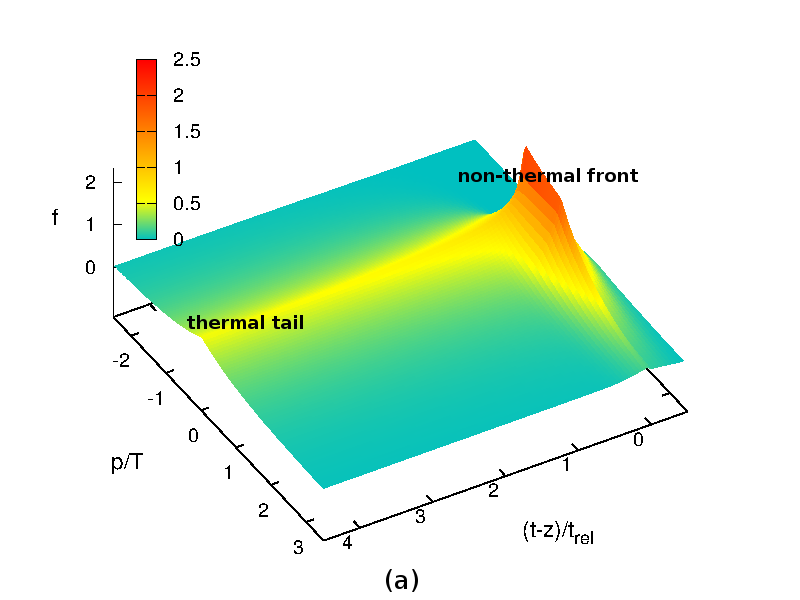}}
 {\includegraphics[width=0.45\textwidth]{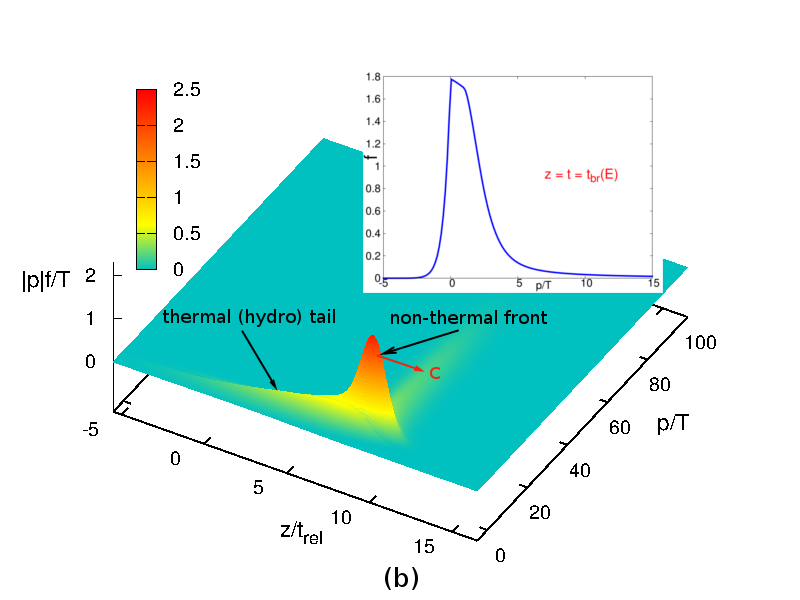}}
 \end{center}
\caption{Charateristic ``front+tail'' struction of the gluon distribution: (a) the analytical solution with the steady source \eqref{steady} and $p_0 = T$; (b) the numerical solution to Eq. \eqref{eqL} with $E=90~T$ and $t=\tbr=9.5~\trel$.}\label{fig:source}
\end{figure}

A high-energy jet typically has the branching time $\tbr\gg \trel$. And the energy flux carried away by the radiated gluons is a slowly varying function of time within a time interval $\Delta t \sim \trel$. Motivated by such an observation, we first consider a simpler problem, where the inelastic collision term in the r.h.s. of \eqref{eqL} 
is replaced by a steady source for particles
with longitudinal momentum $p_0$ which propagate at the speed of light:
\begin{align}\label{steady}
\mcal{C}_{\rm br}[f_\ell]\,\to\,T\delta(p-p_0)\delta(z-t)\qquad\text{with}\qquad p_0\sim T.
\end{align}
We find that analytical solutions exist in this case\cite{Iancu:2015uja}, which is illustrated in Fig.~\ref{fig:source}~(a). In this figure, one sees a two component structure, with a front and a tail. The {\em front} is made with
gluons travelling at the speed of light. These gluons are injected by the source at $z\gtrsim t-\trel$ before $t$ and do not have enough time to establish local thermalization. The {\em tail} lies behind the front. At $z\lesssim t-\trel$, it is made of gluons which have relaxed to local thermal equilibrium with the QGP. This picture can be justified by the Green function for this problem, which takes the form\cite{Iancu:2015uja}
\beq\label{MB}
f_G\simeq \frac{\rme^{-|p|/T}}{2}\,
 \frac{\rme^{-\f{\left(z- p_0 \trel/T\right)^2}{4 t}}}{2 \sqrt{\pi  t/\trel}}\qquad\text{at }t\gtrsim \frac{t_{rel}p_0}{T}\text{ and } \trel.
\eeq
It also tells us that these gluons, together with the background QGP, behave hydrodynamically with the (longitudinal) size of disturbance $\sqrt{\bra (z-\frac{t_{rel} p_0}{T})^2\ket}\propto \sqrt{t}$. Numerical solutions to \eqref{eqL} also show the ``front+tail'' structure (See Fig.~\ref{fig:source}~(b) for example). Therefore, the analytical solution in this simplified problem captures the generic feature of jet evolution before $t\sim \tbr$. Our calculations support that the motion of these gluons in the tail can be described by hydrodynamics\cite{He:2015pra}.
 
\begin{figure}[h]
\begin{center}
\includegraphics[width=.45\textwidth]{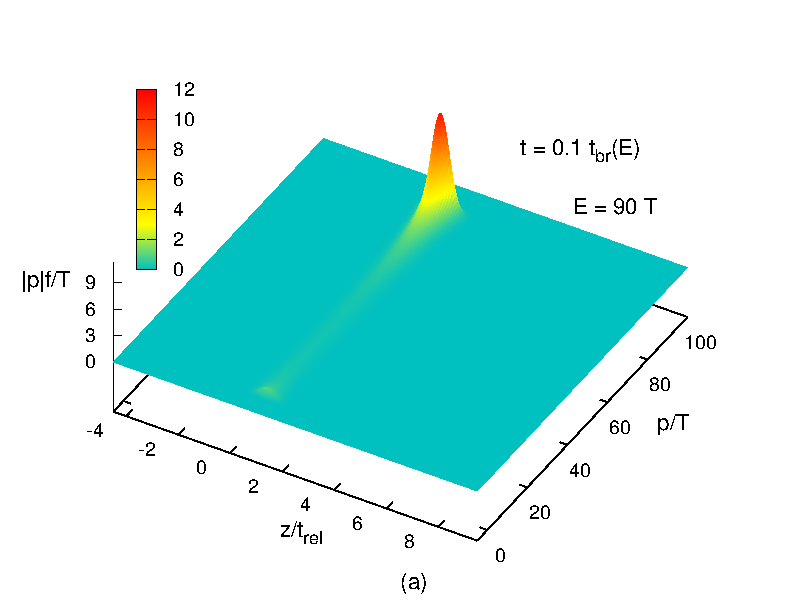}
\includegraphics[width=.45\textwidth]{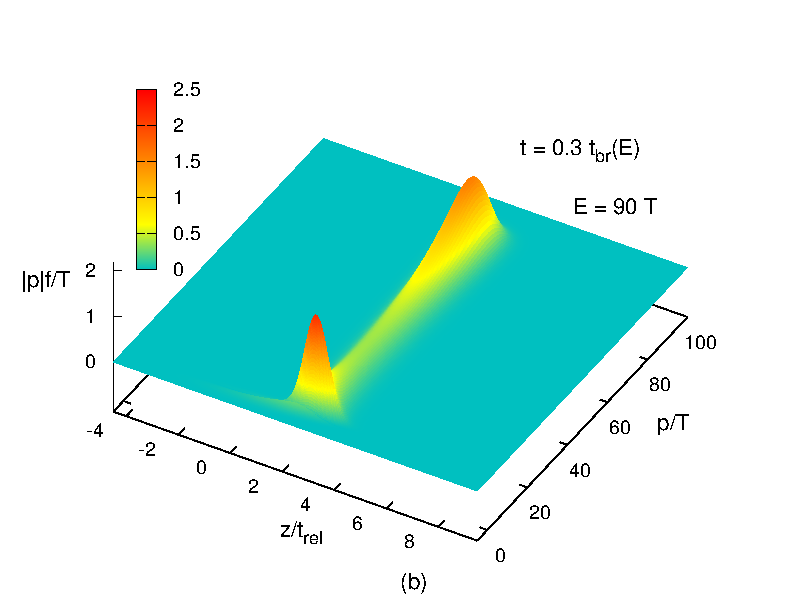}
\end{center}
\caption{The phase-space energy distribution $|p|f_{\ell}/T$ produced by an energetic
jet with $E=90\,T$ at two successive times: (a) an early time $t=0.1\tbr(E)$, when the jet is almost
unquenched; (b) a larger time $t=0.3\tbr(E)$, when the jet is partially quenched.}
\label{fig:early}
\end{figure}

We now turn to the solutions to the general equation \eqref{eqL} with the initial condition \eqref{init}. The time scales inherent in this equation are the branching time $\tbr(E)$ for the LP
%that would set the lifetime of the LP (and hence of the jet) 
%in a sufficiently large medium ($L >\tbr(E)$), 
and the relaxation time $\trel$. We shall take $E=90~T$ for example to discuss the features of jet evolution. At $t\gtrsim 1.5$, the jet is found to be fully quenched. The LP disappears completely and almost all the gluons resulted from multiple branching of the jet establish local thermal equilibrium with the plasma. At the LHC, one has $L<\tbr(E)$, as already mentioned,
hence the LP is expected to survive in the final state. This is indeed visible in the numerical results
displayed in Fig.~\ref{fig:early}, as numerically obtained for $E=90\,T$ 
and $\tbr(E)= 9.5\,\trel$ \cite{Iancu:2015uja}.

Fig.~\ref{fig:early} (a) shows the distribution at the very early time $t=0.1\tbr(E)$,
when most of the energy is still carried by the LP. Hence the energy distribution $|p|f_{\ell}/T$
shows a pronounced peak at $p/T=90$ and at $z=t$. Yet, this peak shows some spreading in
$p$, as a consequence of early emissions, which are necessarily soft: the typical quanta emitted
up to time $t$ have $p\lesssim \omega_{\rm br}(t)=
\alpha_s^2\hat q t^2$. 

At the larger time $t=0.3\tbr(E)$, cf. Fig.~\ref{fig:early} (b), the softening
of the distribution in $p$ is clearly visible, albeit a pronounced LP peak still exists. One can
now distinguish the characteristic ``front+tail'' structure anticipated in Fig.~\ref{fig:source}.
The front at $z=t$ involves relatively hard gluons, whose momentum distribution (within
the range $T < p \ll E$) is given by the scaling spectrum $f_{\ell}\propto
1/p^{3/2}$, as expected for the LPM spectrum  
\cite{Baier:1996kr, Zakharov:1996fv, Baier:2000sb,Blaizot:2013hx}. Similar results have been obtained in a kinetic theory study of the thermalization 
of the background quark-gluon plasma \cite{Kurkela:2014tea}.

The front in Fig.~\ref{fig:early} (b) also shows 
a secondary peak at $p=T$, due to the accumulation of gluons
at the lower end of the cascade. Such gluons are abundantly produced via branchings and they
cannot thermalize instantaneously --- rather, they need a time $\sim\trel$ to that aim.
Yet, since $t=0.3\tbr(E)\simeq 3\trel$ is relatively large compared to $\trel$, a thermalized tail
at $z\lesssim t-\trel$ develops indeed, as visible too in Fig.~\ref{fig:early}.b.
This tail carries the energy lost by the jet towards the medium.
%
%\begin{figure}[t]
%\begin{center}
%\includegraphics[width=0.35\textwidth]{Iancu_ftzE90}
%\caption{The gluon distribution in $p$ at $t=z$ for $E=90\,T$ and for 4 values of time.
%The figures show a rather broad window of approximate scaling behavior, $f_{\ell}\propto {1}/{p^{3/2}}$,
%at not too large times $t\lesssim \tbr(E)$.
%}\label{fig:deviScal}
%\end{center}
%\end{figure}

%\begin{figure}[h]
%\begin{center}
%\includegraphics[height=0.2\textheight]{e3E90discuss}
%\hspace{2mm}
%\includegraphics[height=0.2\textheight]{pTbalancecentral}
%%\includegraphics[width=.45\textwidth]{e3E90.png}
%\end{center}
%\caption{The phase-space energy distribution $|p|f_{\ell}/T$ produced by an energetic
%jet with $E=90\,T$ at two successive times: (a) an early time $t=0.1\tbr(E)$, when the jet is almost
%unquenched; (b) a larger time $t=0.3\tbr(E)$, when the jet is partially quenched.}
%\label{fig:early}
%\end{figure}

These results allow for a qualitative comparison with the phenomenology of di-jet asymmetry 
at the LHC  \cite{Aad:2010bu,Chatrchyan:2011sx}. (For a more quantitative comparison,
one could take $T=0.5\,{\rm GeV}$ and %, $\hat q=1\,{\rm GeV}^2/{\rm fm}$, and
$\trel=1\,{\rm fm}$.)
 The early situation in Fig.~\ref{fig:early} (a), where the jet is essentially
 unquenched, is illustrative for the leading jet, which crosses at most a very narrow slab of matter. 
 The situation in Fig.~\ref{fig:early} (b), where the jet looks partially quenched,
is representative for the subleading jet in a di-jet event characterized 
by a large asymmetry.  For even larger times, $t\gtrsim \tbr(E)$,
both the LP and the front would disappear and the whole energy would be found in the thermalized tail
  \cite{Iancu:2015uja}.

%\bibliographystyle{elsarticle-num}
%\bibliography{refs}
%\end{document}






% \end{thebibliography}

\end{document}